\documentstyle[12pt,a4,amssymb,psfig]{article}

\newcommand{\Rreal}{{\Bbb R}}
\newcommand{\lagrangiano}{{\cal L}}
\newcommand{\be}{\begin{equation}}
\newcommand{\ee}{\end{equation}}
\newcommand{\bea}{\begin{eqnarray}}
\newcommand{\eea}{\end{eqnarray}}

\begin{document}

\begin{titlepage}

\begin{flushright}
{\tt FTUV/93-34\\
     IFIC/93-34\\
     hep-th/9312030}
\end{flushright}

\vfill

\begin{center}

{\bf{\Large Quantum Cosmological Approach to
	    2d Dilaton Gravity}\footnote[2]{Work
partially supported by the {\it Comisi\'on Interministerial de Ciencia y
Tecnolog\'{\i}a} and {\it DGICYT}.}
}

\bigskip
\bigskip
\bigskip

	J. Navarro-Salas and C. F. Talavera\\

\bigskip

\begin{center}
	 Departamento de F\'{\i}sica Te\'orica and\\
	IFIC, Centro Mixto Universidad de Valencia-CSIC.\\
	Facultad de F\'{\i}sica, Universidad de Valencia,\\
	Burjassot-46100, Valencia, Spain.
\end{center}
\bigskip
\today
\bigskip

\end{center}

\begin{center}
{\bf Abstract}
\end{center}

	We study the canonical quantization of the induced 2d-gravity
and the pure gravity CGHS-model on a closed spatial section.
The Wheeler-DeWitt equations are solved in
(spatially homogeneous) choices of
the internal time variable and the space of solutions
is properly truncated
to provide the physical Hilbert space.
We establish the quantum equivalence of both models and
relate the results with the covariant phase-space quantization.
We also
discuss the relation between the quantum wavefunctions and the
classical space-time solutions and propose the wave function
representing the ground state.

\vfill

\end{titlepage}
\newpage

\section{Introduction}

	One of the most recent research directions in quantum gravity is
the study of two dimensional dilaton models. Lower-dimensional
gravity provides useful
toy models for understanding the quantum mechanics of
generally covariant theories.
In two space-time dimensions, the Hilbert-Einstein action is trivial but
the constant curvature condition can serve as the natural analogue of
the vacuum Einstein equations \cite{Jack,Teit}.
However, this equation cannot be derived from a local generally
covariant action unless a scalar field $\Phi$ is introduced in the
theory. The easiest way to obtain this equation from a local action
is when the scalar field $\Phi$ plays the role of a lagrangian
multiplier \cite{Jack,Teit}.

        The constant curvature equation can also be obtained from
the induced 2d-gravity action \cite{Poly}
\be
\label{indu}
S = {c \over 96\pi} \int d^2x \sqrt{-g} \left( R \square^{-1} R +
	4\lambda^2 \right) \, .
\ee
This non-local action
(induced by massless matter fields \cite{Poly})
can be converted into a local one by introducing
a dilaton scalar field $\Phi$ (we omit the coupling constant)
\be
\label{indulocal}
S = {1 \over 2 \pi} \int d^2 x \sqrt{-g} \left[ (\nabla\Phi)^2 +
	2 R \Phi + 4 \lambda^2 \right] \, .
\ee

        Recently, another manageable and intriguing model of two-dimensional
gravity has been proposed by
Callan, Gidding, Harvey and Strominger \cite{CGHS} with the crucial property
of containing black holes. This model
provides an excellent scenario to study black hole evaporation and
back-reaction in a full quantum gravity setting (for a review see
\cite{HarvStro}).
The classical action of the CGHS model (gravity coupled to a
dilaton field $\Phi$ and conformal matter fields $f_i$,
$i=1,\ldots,N$) is given by
\be
\label{accghs}
S = {1 \over 2 \pi} \int d^2x \sqrt{-g} \left[ e^{-2\Phi} ( R +
	4 (\nabla\Phi)^2 + 4 \lambda^2 ) -
	\sum_{i=1}^{N} {1\over2} (\nabla f_i)^2
	\right] \, .
\ee
In a canonical framework the two-dimensional
space-time manifold is usually required
to have the topology of $\Sigma\times{\Rreal}$, where the spatial section
$\Sigma$ can be either ${\Rreal}$ or $S^1$. For the non-compact case the
classical solutions of (\ref{accghs})
admit the standard black hole interpretation.
When the spatial section is compact it is not clear whether the
black hole interpretation can be maintained \cite{Miko}. However, in the
latter case, and when no matter fields are present, the model
--and the induced 2d-gravity (\ref{indulocal}) as well-- has the
remarkable property of having  all the classical solutions
spatially homogeneous. This is just the main ingredient of the
minisuperspace approach in quantum cosmology \cite{DeWitt,Hartle}
(see also the review \cite{Halli}).
Nevertheless, we should stress
that, in contrast to $3+1$-dimensional gravity, where the
minisuperspace models are approximations to the full theory,
in the above models
the condition of having spatially homogeneous solutions comes from
the field equations themselves.
Therefore, the minisuperspace approach can be an exact way
to quantize the pure gravity 2d-dilaton models \cite{Pepeii}.

	In this paper we shall study, in a parallel way, the
induced 2d-gravity (\ref{indulocal}) and the pure gravity
CGHS-model for a closed spatial section in an exact canonical setting.
We shall consider the ADM formulation of the 2d-dilaton models
in a natural choice of the time slicing. The gauge-fixing will be
spatially homogeneous and together with the diffeomorphism constraint
reduces the theories to a finite number of degrees of freedom. This
strategy allows us to go fairly far in solving the models.
Other approaches to the quantization of 2d-gravity models
(mainly using Dirac operator methods and BRST techniques)
can be seen in \cite{Isler}. In these works the supermomentum
constraint is imposed at the quantum level. The possible
equivalence with
our results is an interesting and
non-trivial point, but it is out of the scope of the
present paper.

	In section 2 we present the classical solutions of the
induced 2d-gravity and the CGHS-model focusing on their
covariant phase-space structure. We shall show in this way
that both models have the same reduced phase space.
In section 3 we develop the ADM formulation of the models in a generic
spatially homogeneous gauge. This enables us to reduce the theory to
a finite number of degrees of freedom and solve the (reduced)
Wheeler-DeWitt equation exactly. The analysis of the quantum solutions
of the induced 2d-gravity suggests a natural choice of the time
slicing as well as a canonical transformation of the
``minisuperspace'' variables of the CGHS-model.
The problem of the Hilbert space is considered in section 4.
The space of
solutions to the Wheeler-DeWitt equation is truncated to provide
the proper Hilbert space. We shall also establish a quantum equivalence
between both models in agreement with the equivalence predicted by the
covariant phase-space quantization. In section 5, and based on the
particular choice of time introduced previously, we discuss the
classical behaviour of the quantum wave functions and also propose
the wave function representing the ground-state.

\section{Classical solutions and covariant phase space}
\label{s:classical}

\subsection{CGHS-model}

	In the pure gravity CGHS-model the classical solutions
for the metric and the dilaton field are given, in the conformal gauge, by
\bea
e^{-2\Phi} &=& -\lambda^2 p m + {M \over \lambda} \, ,\\
ds^2 &=& { \partial_+ p \partial_- m \over -\lambda^2 p m + {M\over\lambda}}
	( -dt^2 + dx^2) \, ,
\eea
where $p$ ($m$) is a function of $x^+$ ($x^-$) ($x^\pm = t \pm x$),
$\lambda = + \sqrt{\lambda^2}$ and
$M$ is a constant parameter playing the role of the black hole mass
when $\Sigma = {\Rreal}$. When the theory is defined
on a closed spatial section,
$\Sigma = S^1$, the requirement of periodicity of the metric and the
dilaton implies the following monodromy
transformations of the functions $p$ and $m$:
\bea
p(x^+ + 2\pi) &=& e^r p(x^+) \, , \label{monoi}\\
m(x^- - 2\pi) &=& e^{-r} m(x^-) \, ,\label{monoii}
\eea
where $r$ is an arbitrary parameter. Despite of the appearance of
arbitrary functions in the general solutions, the space of non-equivalent
solutions --under space-time diffeomor\-phisms-- is finite-dimensional.
This fact can be accomplished by evaluating the symplectic structure
of the model
\be \label{dosviii}
\omega = \int_0^{2\pi} dx (-\delta j^0) \, ,
\ee
where $\delta$ stands for the exterior derivative on the space of
classical solutions (i.e., the so-called covariant phase space
\cite{Crnko}), and $j^\mu$ is the symplectic current potential defined,
in general, as \cite{Pepe}
\be
\delta{\lagrangiano} = \partial_\mu j^\mu + {\delta S \over \delta\varphi}
	\delta\varphi \, ,
\ee
where $S=S(\varphi)$ is the action functional.

        In the present case the two-form $\omega^0 = -\delta j^0$ turns
out to be a total derivative reflecting the absence of local degrees
of freedom. More precisely, $\omega^0 = \partial_x W$, where
\be
W = \delta\left({M \over \lambda}\right) \delta\ln\partial_- m +
	\delta(-\lambda^2 p m ) \delta\ln(p \partial_- m) \, .
\ee
The symplectic form (\ref{dosviii}) then becomes
\be
\omega = W(x+2\pi) - W(x) \, ,
\ee
and, using (\ref{monoi}) and (\ref{monoii}), one find that any dependence
with respect to the $p$ and $m$ functions drops out and we are left with
(see ref. \cite{Miguel})
\be \label{omega}
\omega = -\delta{M\over\lambda} \delta r \, .
\ee
Therefore, the simple expression (\ref{omega}) captures, in an appealing
way, the canonical structure of the model. We can construct
explicitly the set of non-equivalent solutions by choosing appropriately
the functions $p$ and $m$ verifying the monodromy condition (\ref{monoi})
and (\ref{monoii}). It is not difficult to arrive at the following
expressions (see also \cite{Mishima})
\bea
ds^2 &=& {1\over |\lambda|^2} \left({r\over 2\pi}\right)^2
        {e^{{r\over\pi} t} \over
                {M\over\lambda} - s e^{{r\over\pi} t}} (-dt^2 +dx^2)
	\, , \label{dosx} \\
e^{-2\Phi} &=& {M\over\lambda} - s e^{{r\over\pi} t}
	\, , \label{dosxi}
\eea
where $s$ refers to the sign of the cosmological constant $\lambda^2$.
In this way one immediately finds that the
covariant phase space (i.e. the reduced phase space)
corresponds to one single degree of freedom.

When $M=0$ and the cosmological constant is negative the solution
becomes the standard ``linear dilaton'' vacuum solution:
\bea
ds^2 &=& -dt^2 + dx^2 \, ,\label{dosxiaa} \\
\Phi &=& -\lambda t   \, . \label{dosxiab}
\eea

\subsection{Induced 2d-gravity}

        Here we present the non-equivalent classical solutions of
the induced 2d-gravity (\ref{indulocal}) on the cylinder (for a
detailed discussion see reference \cite{Pepe}):
\bea
ds^2 &=& {4\over|\lambda|^2} \left({r \over 2\pi}\right)^2
	{e^{{r\over\pi}t} \over \left(1-s e^{{r\over\pi}t}\right)^2}
	(-dt^2+dx^2) \, , \label{dosxv} \\
\Phi &=& \ln \alpha{ \left( 1-s e^{{r\over\pi}t}\right)^2 \over
        4\left( \sinh{r\over4\pi} \right)^2 } \, , \label{dosxvi}
\eea
where $\alpha$ is an arbitrary positive constant.
The solutions (\ref{dosxv}) of the Liouville equation involve the hyperbolic
and parabolic monodromy matrices only. The elliptic monodromy matrices are
forbidden by the additional equations of motion. When the cosmological
constant is negative only the hyperbolic monodromies
$\left( \begin{array}{cc}
        e^r & 0 \\
        0   & e^{-r}
        \end{array} \right)$
are allowed and, for $\lambda^2 > 0$, we also have the parabolic solution
$\left( \begin{array}{cc}
        1 & b \\
        0 & 1
        \end{array} \right)$
(independent of the `b' parameter) obtained as the limit $r\rightarrow 0$ of
(\ref{dosxv}), (\ref{dosxvi}),
\bea
ds^2 &=& {1 \over 2|\lambda|^2} {1\over t^2} (-dt^2 + dx^2) \, , \\
\Phi &=& \log \alpha{t^2 \over \pi^2} \, .
\eea
The symplectic structure of the model reads as \cite{Pepe}
\be
\omega = 4 \delta \ln\alpha \delta r \, .
\ee

	As a result of the above we can conclude that both
models have the same covariant phase space. It is the
cotangent bundle of two disconnected ${\Rreal}^{+}({\Rreal})$
spaces for the induced 2d-gravity when $\lambda^2<0$ ($\lambda^2>0$)
and the pure gravity CGHS-model if $\lambda^2>0$ ($\lambda^2<0$).
The two sectors correspond to whether the dilaton field is expanding
or contracting.
In quantizing these symplectic manifolds
the quantum states are represented by square integrable
functions depending on the ``configuration'' constant of motion.
We shall see in the next section how this prediction for the
Hilbert space is consistent with the quantization coming from
the Wheeler-DeWitt equation.

\section{The Wheeler-DeWitt equation}

\subsection{Induced 2d-gravity}

        To study the 2d-dilaton gravity from a canonical
point of view we first present the ADM formulation \cite{ADM}.
We parametrize the two-dimensional metric as
\be
g_{\mu\nu} = \left( \begin{array}{cc} -N^2 + N_1 N^1 & N_1 \\
                                           N_1       & a^2
                    \end{array}
             \right) \, ,
\label{btresi}
\ee
where $N$ and $N^1$ are the lapse and shift functions and $a^2$
plays the role of the ``spatial metric''. Using the two-dimensional
identity $\sqrt{-g} R = -2 \partial_t (a K) + 2(a (KN^1 - a^{-2}N'))'$,
where $K$ is the extrinsic curvature scalar
($K = {1\over a^2 N} (N_{1|1} - a \dot{a} )$),
it is straightforward
to rewrite (\ref{indulocal}) in the hamiltonian form
(see \cite{Torre} for an earlier study):
\be
S = \int d^2x (\pi_a \dot{a} + \pi_\Phi \dot{\Phi} - N {\cal C} -
N^1 {\cal C}_1 ) \, .
\label{btresii}
\ee
The canonical momenta are
\bea
\pi_a &=& {4\over N} ( \Phi' N^1 - \dot{\Phi} ) \, , \label{btresiii}\\
\pi_\Phi &=& {2 a \over N} (\Phi' N^1 - \dot{\Phi}) + {4\over N}
        \left( (aN^1)' - \dot{a} \right) \, , \label{btresiv}
\eea
and the functions
\bea
{\cal C}_1 &=& \Phi' \pi_\Phi - \pi'_a a \, \label{btresv} \\
{\cal C} &=& {1\over 16} a \pi_a^2 - {1\over4} \pi_a \pi_\Phi -
        4 a \lambda^2 -{1\over a} \Phi'^2 + 4(a\Phi')' \label{btresvi}
\eea
are the supermomentum and hamiltonian constraints respectively.

        To solve the theory we shall first
reduce it removing the non-dynamical
variables by gauge fixing. The canonical form of the model
(\ref{btresii})-(\ref{btresvi}) suggests to choose the ``internal'' time
variable as (for a review on the problem of time see \cite{Isham})
\be
{\cal T}(\Phi,a) = \chi(t) \, ,\label{btresvii}
\ee
where $\chi$ is an arbitrary function. Furthermore, we can also exploit
the $\Sigma=S^1$-diffeo\-mor\-phism invariance of the theory to fix
the space coordinate in such a way that
\be
a = a(t) \, . \label{btresviii}
\ee
The above gauge conditions together with the constraint (\ref{btresv})
imply that
\bea
\Phi &=& \Phi(t) \, , \label{btresix} \\
\pi_a &=& \pi_a(t) \, , \label{btresx}
\eea
and then the hamiltonian constraint reduces to (see also \cite{Pepeii})
\be
{\cal C} = {1 \over 16} a \pi_a^2 - {1\over4}\pi_a\pi_\Phi -
        4 a \lambda^2 \, , \label{btresxi}
\ee
(now $\pi_\Phi$ stands for $\pi_\Phi(t) \equiv \int dx \pi_\Phi(t,x)$ ).
If we choose particular functions $\chi$ and ${\cal T}$
in (\ref{btresvii}) it is
still possible to further reduce the theory. However we prefer to maintain
the function $\chi$ undetermined and go to the quantum theory by using
the reduced hamiltonian constraint (\ref{btresxi}).

	The quantum mechanics of the model is governed by the operator
version of the classical constraint ${\cal C} = 0$, i.e., the
Wheeler-DeWitt equation. Our task is now to propose a Wheeler-DeWitt
equation incorporating the usual factor ordering ambiguities and
the restriction $a(t)>0$ of the scale variable. The latter point can be
incorporated in a natural way by choosing the affine variables $a$,
$p_a \equiv a\pi_a$ as the basic ones for the quantization \cite{Klauder}.
The reason is that the operator $\hat{\pi}_a = -i \hbar
{\partial \over \partial a}$ fails to be self-adjoint in $L^2({\Rreal}^+,
da)$, but the operators $\hat{a} = a$, $\hat{p}_a = -i\hbar a
{\partial \over \partial a}$ are self-adjoint in the space
$L^2({\Rreal}^+, {da\over a} )$.
In terms of the classical affine
variables, the constraint ${\cal C}=0$ reads
\be \label{tresviii}
{\cal C} = {1\over 16a} p_a^2 -
        {1\over 4a} p_a \pi_\Phi - 4\lambda^2 a \, ,
\ee
and the quantum constraint $\hat{{\cal C}}$ can be written
as (self-adjoint with respect to the measure ${da\over a}d\Phi$)
\bea \label{wdweq}
\hat{{\cal C}}&=&\Bigl[
 -{1\over16} \hat{a}^{\alpha+i\beta} \hat{p}_a \hat{a}^{-1-2\alpha} \hat{p}_a
\hat{a}^{\alpha-i\beta} +  \nonumber \\
& & \hphantom{\Bigl[}
 {1\over8} \left( \hat{a}^{\gamma+i\sigma} \hat{p}_a
\hat{a}^{-\gamma-i\sigma-1} + \hat{a}^{\gamma+i\sigma-1} \hat{p}_a
\hat{a}^{-\gamma-i\sigma} \right) \hat{\pi}_\Phi +
 4 \lambda^2 \hat{a} \Bigr]   \, ,
\eea
where $\alpha$, $\beta$, $\gamma$ and $\sigma$ are
arbitrary factor-ordering parameters.

	To solve the Wheeler-DeWitt equation,
$\hat{\cal C}\Psi = 0$,
we shall make use of the classical constant of motion
$\pi_\Phi$. It is interesting to identify $\pi_\Phi$ on the covariant
phase space.
Inserting the classical solutions (\ref{dosx}) and (\ref{dosxi}) into
(\ref{btresiv}) we can identify the constant of motion on the covariant
phase space as
\be
\pi_\Phi = -{r\over\pi} \, .
\label{btresxiv}
\ee

Expanding the wave function $\Psi$ in $\hat{\pi}_\Phi$
eigenstates
\be
\Psi = \int dq e^{{iq\Phi\over\hbar}}
\Psi_q(a) \, ,
\label{tresxi}
\ee
the Wheeler-DeWitt equation separates, and the functions $\Psi_q$
satisfy
\be \label{tresxii}
\left(
{d^2\over dz^2} +
{1\over z}(1 - 2\zeta){d\over dz} +
{1\over z^2}(\zeta^2-\nu^2) +
16 {\lambda^2\over\hbar^2}
\right) \Psi_q(z) = 0 \, ,
\ee
where $z\equiv a$ and
\bea
\zeta &=& {1\over2} \left( 1 + 4{iq\over\hbar} + 2i\gamma \right)
        \, , \label{tresxiia} \\
\nu^2 &=& {1\over4} \left( 1- 16{q^2\over\hbar^2} - 8\gamma^2 +
4\alpha(\alpha+1) + 16 {q\over\hbar}(\sigma-\gamma) \right)
        \, . \label{tresxiib}
\eea

	The solutions of the above equation are
\be
\Psi_q(z) = z^\zeta {\cal Z}_\nu (kz) \, ,
\ee
where
\be
k = 4 {\lambda\over\hbar} \, , \label{tresxiv} \\
\ee
and ${\cal Z}_\nu (kz)$ are ordinary (modified) Bessel functions for
$\lambda^2 >0$ ($\lambda^2<0$) with order $\nu$.

	We should remark now that, irrespective of normal ordering
prescription of (\ref{wdweq}), the orders of the Bessel functions
entering in the general solutions of the Wheeler-DeWitt equation
are always purely real or imaginary.

        To finish our analysis of the Wheeler-DeWitt equation
we shall relate it with the so-called reduced phase space approach.
This will be of interest by itself but in our case we shall exploit
it later to propose the wave function of the ``ground state''.
The main point is to reduce the theory to a genuine canonical form
before going to the quantum theory. This requires choosing  a particular
function ${\cal T}$ in the time-fixing condition ${\cal T}(\Phi,a) = t$
(we have set $\chi = t$ in (\ref{btresvii})). The momentum conjugate to
${\cal T}$ expressed in terms of ${\cal T}$ and the independent canonical
variables --satisfying the constraint ${\cal C} = 0$-- plays the role of
the hamiltonian ($H_{red}$) for the reduced phase space.

        There are various useful definitions of internal time in the
literature (see the review \cite{Isham}). In the light of expression
(\ref{tresxi}) a choice of time which naturally arises in the
induced 2d-gravity is given in terms of the dilaton.
Other (spatially homogeneous)
choices of time in 2d-dilaton gravity have been considered in
\cite{Banks}.
The choice $\Phi={\cal T}$ yields to the following reduced hamiltonian
($H_{red} = -\pi_\Phi$)
\be
H_{red} = -{1\over4} p_a - 4 \lambda^2 {a^2\over p_a} \, ,
\label{btresxxii}
\ee
which has the property of being time-independent.
The reduced degrees of freedom could then be quantized in the conventional
way through the Schr\"odinger equation. Ignoring operator ordering ambiguities,
the Schr\"odinger equation can be converted, in our case, into the
Wheeler-DeWitt equation.

\subsection{CGHS-model}

	Using the ``homogeneous'' gauge  (\ref{btresvii}),
(\ref{btresviii}), and arguing as in the induced 2d-gravity case, we
can reduce the pure gravity CGHS-model to the mechanical action
\be
\int dt \left[ 	\pi_\Phi \dot{\Phi} +
		\pi_a \dot{a} -
		N {\cal C}
	\right] \, ,
\label{ctresxxii}
\ee
for which the hamiltonian constraint is given by
\be
{\cal C} = {1\over4} a \pi_a^2 e^{2\Phi} +
	{1\over4} \pi_\Phi \pi_a e^{2\Phi} -
	4 a \lambda^2 e^{-2\Phi}
\, .
\label{ctresxxiii}
\ee
For this model the analogue of the linear constant of motion (\ref{btresxiv})
is now
\be
{1\over2} \pi_\Phi + a \pi_a
\, .
\label{ctresxxiv}
\ee
On the covariant phase space we can identify (\ref{ctresxxiv}) as
\be
{M \over \pi \lambda} r
\, .
\label{ctresxxv}
\ee
The constant of motion (\ref{ctresxxiv}) suggests to introduce a new
set of ``minisuperspace'' variables
\bea
T &=& 2 \Phi + \ln a \, , \label{ctresxxvia}\\
A &=& a e^{-2\Phi} \,. \label{ctresxxvib}
\eea
These variables seem to be the natural ones to study the model
since lead, for the time-choice ${\cal T} = T$, to a time-independent
reduced Hamiltonian ($H_{red} = -\pi_T = -{1\over2} ( a \pi_a + {1\over2}
\pi_\Phi )$)
\be
H_{red} = {1\over3} \left[ A \pi_A \pm 2 A \sqrt{\pi_A^2 + 12 \lambda^2}
\right] \, .
\label{ctresxxvii}
\ee
In terms of the new variables the function ${\cal C}$ is
given by ($p_A \equiv A \pi_A$)
\be
{\cal C} = -{1\over 4 A} p_A^2 +
	{1 \over 2 A} p_A \pi_T +
	{3 \over 4 A} \pi_T^2 -
	4 A \lambda^2
\, ,
\label{ctresxxviii}
\ee
and the quantum counterpart of the classical constraint can be written as
\bea
& &\Bigl[
{1\over4} \hat{A}^{\alpha+i\beta} \hat{p}_A \hat{A}^{1-2\alpha} \hat{p}_A
\hat{A}^{\alpha-i\beta} +  \nonumber \\
& & \hphantom{\Bigl[}
{1\over4} \left( \hat{A}^{\gamma+i\sigma} \hat{p}_A
\hat{A}^{-\gamma-i\sigma-1} + \hat{A}^{\gamma+i\sigma-1} \hat{p}_A
\hat{A}^{-\gamma-i\sigma} \right) \hat{\pi}_T +
{3\over4} \hat{A}^{-1} \hat{\pi}_T^2 -
4 \hat{A} \lambda^2 \Bigr] \Psi = 0 \, ,\quad
\label{ctresxxix}
\eea
where $\alpha, \beta, \gamma$ and $\sigma$ are arbitrary factor-ordering
parameters.
Expanding the wave functions in
$\hat{\pi}_T$ eigenstates
\be
\Psi = \int dq e^{i q T \over \hbar} \Psi_q(A)
\, ,
\label{ctresxxx}
\ee
and plugging (\ref{ctresxxx}) into (\ref{ctresxxix}) we also find the
differential equation (\ref{tresxii}) for the function $\Psi_q(z\equiv A)$,
where the parameters $\zeta$ and $\nu^2$ are now given by
\bea
\zeta &=& {1\over2} \left[ 1 + 2i {q\over\hbar} + 2 i \beta \right] \,
\label{ctresxxxi} \\
\nu^2 &=& {1\over4} \left[ 1 - 16 {q^2 \over \hbar^2} +
	4 \alpha (\alpha+1) + 8 {q\over\hbar} (\sigma-\beta) \right ] \, .
\label{ctresxxxii}
\eea
As for the induced 2d-gravity, the general solution of the above equation
is given in terms of different kind of Bessel functions:
$\Psi_q = z^\zeta {\cal Z}_\nu (k z)$, where $k$ is also given by
(\ref{tresxiv}) and $z\equiv A$
but now ${\cal Z}_\nu (kz)$ is a modified (ordinary)
Bessel function for $\lambda^2 > 0$ ($\lambda^2 < 0$).

\section{Hilbert space}

	To construct a suitable Wheeler-DeWitt equation we have
required hermiticity of the Wheeler-DeWitt operator $\hat{\cal C}$ with
respect to the standard inner product
\be \label{cuatroi}
<\Psi_1|\Psi_2> = \int {da\over a} d\Phi \Psi_1^* \Psi_2 \, ,
\ee
where ${da\over a} d\Phi$ is the volume element of minisuperspace.
In canonical quantum gravity the use of (\ref{cuatroi}) as the
physical scalar product for the solutions of the Wheeler-DeWitt
equation is rather problematic \cite{Pepeii}. In (\ref{cuatroi}) we
are indeed integrating over the reduced minisuperspace
configuration variable and
some sort of ``time'' variable as well. From the functional integral
point of view of quantum gravity \cite{Hartle}, the expression
(\ref{cuatroi}) would have a proper meaning in terms of the standard
sum over all histories with the appropriate boundary conditions.
Both schemes lead to expect (\ref{cuatroi}) to be potentially divergent
and, therefore, it is natural to define the proper inner product as
the regularized version of (\ref{cuatroi}). In the following we shall
determine the Hilbert space of
our pure gravity 2d-dilaton models (\ref{indulocal}) and
(\ref{accghs}) in terms of normalizable solutions of the
Wheeler-DeWitt equation.

\subsection{Induced 2d-gravity}

	According to section 3, the general solution to the
Wheeler-DeWitt equation, for $\lambda^2 > 0$, can be expanded as
($Re(\nu)\ge 0$, $Im(\nu)\ge 0$)
\be \label{cuatroii}
\Psi = \int dq a^\zeta
        e^{{iq\Phi\over\hbar}}
        \left( A(q) {\cal J}_\nu (kz) + B(q) {\cal N}_\nu (kz) \right) \, ,
\ee
where $\nu$ is given by (\ref{tresxiib}) and $A(q)$, $B(q)$ are arbitrary
complex functions.
Now we want to evaluate the norm of the wave function (\ref{cuatroii})
with respect to (\ref{cuatroi}).
It is not difficult to arrive at the following
expression ($x=kz$)
\bea \label{cuatroiii}
<\Psi|\Psi> &=& {2\pi\hbar \over k}
	\int_{-\infty}^{+\infty} dq \int_0^{+\infty} dx
  \Bigl(
        |A(q)|^2 |{\cal J}_\nu (x)|^2 +
        |B(q)|^2 |{\cal N}_\nu (x)|^2 +
\nonumber \\
 &&
\hphantom{ \pi\hbar \int_{-\infty}^{+\infty} dq }
A^*(q) B(q) {\cal J}_\nu^*(x) {\cal N}_\nu(x) +
        A(q) B^*(q) {\cal J}_\nu(x){\cal N}_\nu^*(x) \Bigr) \, .
\eea

        Now the difficulty is the integration over $x$.
Due to the asymptotic behaviour of the Bessel functions
for large $x$ the integrals of (\ref{cuatroiii}) are divergent.
We can regularize the formal scalar product (\ref{cuatroiii}) by
substituting the integration measure $dx$ in (\ref{cuatroiii}) by
${dx\over x^\epsilon}$ ($\epsilon \gtrsim 0$). The resulting expression
can be easily evaluated and we obtain ($\Theta$ is the step function)
\bea
<\Psi|\Psi> &{\mathop{\sim}\limits_{\scriptscriptstyle\epsilon\to0}}&
        {\hbar \over k} {\Gamma(\epsilon) \over 2^\epsilon}
        \int_{-\infty}^{+\infty} dq \Bigl\{
        \bigl[ \cos(\pi\nu) (|A|^2+|B|^2) + \sin(\pi\nu) (B^* A - A^* B)
        \bigr] \Theta(-\nu^2) + \nonumber \\
        &&
	\hphantom{{\hbar \over k} {\Gamma(\epsilon) \over 2^\epsilon}
        	\int_{-\infty}^{+\infty} dq}
       \bigl[ |A|^2 + |B|^2 \bigr] \Theta(\nu^2) \Bigr\} \, .
\label{cuatroiv}
\eea
The divergent term ${\Gamma(\epsilon)\over 2^\epsilon}$ is an overall
factor and can be eliminated in the definition of the physical scalar
product. The Hilbert space will be made out of normalizable wave functions
respect to the regularized scalar product. We shall return to this point
later.

\bigskip
\bigskip

        Let us analyze now the case of negative cosmological constant.
We can write the general solution to the Wheeler-DeWitt equation as in
(\ref{cuatroii}) but then the ordinary Bessel functions
should be replaced by the modified ones
${\cal I}_\nu$.
The asymptotic behaviour of the combination
${\cal I}_{-\nu} - {\cal I}_\nu$ is different from the corresponding one of
${\cal I}_\nu$. In fact the functions ${\cal K}_\nu =
{\pi\over 2\sin(\nu\pi)} \left( {\cal I}_{-\nu} - {\cal I}_\nu \right)$
decay exponentially for large $x$ but the functions ${\cal I}_\nu$,
instead, grow exponentially. This means that, even if we regularize the
formal scalar product as before, the solutions
${\cal I}_{\nu} + {\cal I}_{-\nu}$ are not normalizable.
Therefore the physical wave functions should be of the form
\be
\Psi = \int dq  e^{{i q \Phi \over \hbar}} a^{{1\over2}+2i{q\over\hbar}}
        C(q) {\cal K}_\nu (4{\lambda\over\hbar}a) \, ,
\label{ccuatrovid}
\ee
where ${\cal K}_\nu$ are the modified Hankel functions.
The point now is to determine the range of variation of the $\nu$ parameter.
If the wave functions are required to be normalizable
in the regularized scalar product we find, for
$\lambda^2 < 0$, that
\be
\nu^2 < {1\over4} \, .
\label{ccuatroviii}
\ee
In the light of expression (\ref{tresxiib}) the natural choice of factor
ordering, leading to the values (\ref{ccuatroviii}) as $q$ varies over
the real line, is given by
\be
\alpha = \beta = \gamma = \sigma = 0 \, .
\label{ccuatroix}
\ee
Therefore the relation between ``$q$'' and the order $\nu^2$ is
\be
\nu^2 = {1\over4} ( 1 - 16 {q^2\over\hbar^2} ) \, .
\label{ccuatrox}
\ee
Note that the solution $q=0$ (i.e. $|\nu|={1\over2}$) is not
normalizable. This definite choice of factor ordering will
allow to establish the equivalence between
the Hilbert space of normalizable solutions to
the Wheeler-DeWitt equation with the Hilbert space predicted
by the covariant phase space quantization (see section (\ref{ss:relation})).

	The consistence between both approaches will also emerge
for $\lambda^2>0$ thus supporting the above choice of factor
ordering. The elementary (normalized) solutions to the Wheeler-DeWitt
equation are
\bea
\pi^{1\over2} {\cal J}_\nu (x) \hfil \; &,& \nu \in [0,{1\over2}] \, ,
\label{gcuatroix}\\
\pi^{1\over2} {\cal N}_\nu (x) \hfil \; &,& \nu \in [0,{1\over2}[ \, ,
\label{gcuatrox}\\
\left( {\pi\over\cosh(\pi Im(\nu))} \right)^{1\over2} {\cal J}_\nu
	\; &,& Re(\nu)=0 \, ,
\label{gcuatroxi}\\
\left( {\pi\over\cosh(\pi Im(\nu))} \right)^{1\over2} {\cal N}_\nu
	\; &,& Re(\nu)=0 \, .
\label{gcuatroxii}
\eea

\subsection{CGHS-model}

        To provide the physical scalar product and the Hilbert space
for the pure gravity CGHS-model we can repeat the same arguments as
for the induced 2d-gravity. The volume element in the minisuperspace
is now ${dA\over A}dT$. For $\lambda^2 < 0$ we find that
the Hilbert space is spanned by
solutions of the form
\be
\Psi(T, A) = \int dq C(q) e^{iqT\over\hbar} A^{{1\over2}+{iq\over\hbar}}
                {\cal K}_\nu (4 {\lambda\over\hbar} A) \, ,
\label{dcuatroxiii}
\ee
where $C(q)$ is a square integrable function vanishing at $q=0$ and
the relation between $q$ and $\nu$ is still given by (\ref{ccuatrox}).
For $\lambda^2 > 0$ the Hilbert space can also be described by
(\ref{gcuatroix}-\ref{gcuatroxii}). The corresponding wave functions
are the same as for the induced 2d-gravity with the replacement
$\Phi \rightarrow T$, $a \rightarrow A$.

	It is clear from the above discussion that a kind of
quantum equivalence between the induced 2d-gravity and the pure
gravity CGHS-model appears in the homogeneous gauge and the
choices of time ${\cal T}=\Phi$, $2\Phi + \ln a$, respectively.
We shall see in section (\ref{ss:relation})
that this equivalence can also be understood in terms of
the covariant phase-space approach.

\subsection{Relation with the covariant phase space}
\label{ss:relation}

	Now we want to discuss the relation between the quantization
carried out in this paper and the
covariant phase space quantization \cite{Pepe,Miguel}.
We know (section \ref{s:classical}) that the reduced phase spaces of
the induced 2d-gravity and the CGHS-model are the same:
$T^*{\Rreal}^+ \cup T^*{\Rreal}^+$ or $T^*{\Rreal} \cup T^*{\Rreal}$
depending on the sign of the cosmological constant.
In general, the quantization of the reduced phase space (symplectic
manifold) requires finding a complete commuting subset of phase space
variables. Due to the cotangent nature of the above phase spaces the
Hilbert space is given by the space of square integrable functions
on the configuration space. Therefore, the Hilbert spaces is either
$L^2({\Rreal}^+) \oplus L^2({\Rreal}^+)$ or
$L^2({\Rreal}) \oplus L^2({\Rreal})$.

	The point now is to see how the above Hilbert spaces
are related with the ones we have obtained in solving the Wheeler-DeWitt
equation. Let us first consider the solutions of the form (\ref{ccuatrovid})
(induced 2d-gravity, $\lambda^2<0$) or (\ref{dcuatroxiii})
(CGHS-model, $\lambda^2>0$) where the ${\cal K}_\nu(x)$ functions are
conveniently redefined as
$\left( {2k\over\hbar}
\Gamma({1\over2}+\nu) \Gamma({1\over2}-\nu) \right)^{1\over2}
{\cal K}_\nu(x)$, ($Re(\nu) < {1\over2}$).
The two sectors of the
Hilbert space can be defined by the natural splitting
$\Psi = \Psi^{(+)} + \Psi^{(-)}$ given by
\bea
\Psi^{(+)} &=& \left( {2k\over\hbar}
		 \Gamma({1\over2}+\nu) \Gamma({1\over2}-\nu)
		\right)^{1\over2}
		\int_0^\infty dq a^{{1\over2}+2i{q\over\hbar}}
		e^{iq{\Phi\over\hbar}} C^{(+)}(q)
		{\cal K}_{|\nu|} (4{\lambda\over\hbar}a)
		\label{fcuatroxiv} \, , \\
\Psi^{(-)} &=& \left( {2k\over\hbar}
		 \Gamma({1\over2}+\nu) \Gamma({1\over2}-\nu)
		\right)^{1\over2}
		\int^0_{-\infty} dq a^{{1\over2}+2i{q\over\hbar}}
		e^{iq{\Phi\over\hbar}} C^{(-)}(q)
		{\cal K}_{|\nu|} (4{\lambda\over\hbar}a)
		\label{fcuatroxv} \, .
\eea
The scalar product turns out to be
\be
< \Psi | \Psi > =
\int_0^\infty dq |C^{(+)}(q)|^2 +
\int^0_{-\infty} dq |C^{(-)}(q)|^2
\label{fcuatroxvi} \, ,
\ee
showing that the Hilbert space obtained from the Wheeler-DeWitt
equation coincides with the one derived from the covariant
phase space: ${\cal H} = L^2({\Rreal}^+,dq) \oplus L^2({\Rreal}^+,dq)$.

	We shall extend our discussion to the case of positive
cosmological constant for the induced 2d-gravity (or, equivalently,
for the CGHS-model when $\lambda^2<0$). The Hilbert space
${\cal H}$ can also be split into two orthogonal subspaces
\be
{\cal H} = {\cal H}^{(+)} \oplus {\cal H}^{(-)} \, \label{fcuatroxvii}
\ee
where ${\cal H}^{(+)}$, ${\cal H}^{(-)}$ are given by solutions
of the form ($Re(\nu) \ge 0$, $Im(\nu) \ge 0$)
\bea
\Psi^{(+)} &=& \left({k\pi\over\hbar}\right)^{1\over2}
		\int_{-\infty}^{+\infty} dq
		a^{{1\over2}+2i{q\over\hbar}} e^{iq{\Phi\over\hbar}}
		A^{(+)}(q)
		\Bigl[ \Theta(q) {\cal J}_\nu (x) + \nonumber \\
		&& \hphantom{\left({k\pi\over\hbar}\right)^{1\over2}
				\int_{-\infty}^{+\infty} dq}
			\Theta(-q) \bigl(
				(\cos(\pi Im(\nu))^{1\over2}
				{\cal N}_\nu(x) +  \nonumber \\
		&& \hphantom{\left({k\pi\over\hbar}\right)^{1\over2}
				\int_{-\infty}^{+\infty} dq}
				\sin(\pi Im(\nu))
				\left(cos(\pi Im(\nu))\right)^{-{1\over2}}
				{\cal J}_\nu (x)
			\bigr)
		\Bigr] \, , \label{fcuatroxviii} \\
\Psi^{(-)} &=& \left({k\pi\over\hbar}\right)^{1\over2}
		\int_{-\infty}^{+\infty} dq
		a^{{1\over2}+2i{q\over\hbar}} e^{iq{\Phi\over\hbar}}
		A^{(-)}(q)
		\Bigl[ \Theta(-q) {\cal J}_\nu (x) + \nonumber \\
		&& \hphantom{\left({k\pi\over\hbar}\right)^{1\over2}
				\int_{-\infty}^{+\infty} dq}
			\Theta(q) \bigl(
				(\cos(\pi Im(\nu))^{1\over2}
				{\cal N}_\nu(x) + \nonumber \\
		&& \hphantom{\left({k\pi\over\hbar}\right)^{1\over2}
				\int_{-\infty}^{+\infty} dq}
				\sin(\pi Im(\nu))
				\left(cos(\pi Im(\nu))\right)^{-{1\over2}}
				{\cal J}_\nu (x)
				\bigr)
		\Bigr] \, . \label{fcuatroxix}
\eea
Using standard integrals of Bessel functions it is not difficult to
check that the (regularized) scalar product is then of the form
($\Psi = \Psi^{(+)} + \Psi^{(-)}$)
\be
< \Psi | \Psi > =
\int_{-\infty}^{+\infty} dq \left( |A^{(+)}(q)|^2 + |A^{(-)}(q)|^2 \right)
\, . \label{fcuatroxx}
\ee
We observe again that the Hilbert space (\ref{fcuatroxvii}-\ref{fcuatroxx})
coincides with the one predicted by the covariant phase-space quantization:
\be
{\cal H}^{(+)} \approx {\cal H}^{(-)} \approx L^2({\Rreal},dq) \, .
\ee

\section{Wave functions and classical space-time}

        In this section we would like to explore the physical meaning
of the quantum wave functions emphasizing their relation with the classical
solutions. First, we have to point out that the parameter
$k = {4\lambda\over\hbar}$ in the argument of the Bessel functions plays
the role of the inverse of the Planck length for these models. Therefore,
for large $x$ ($x \equiv kz \gg 1$) we could expect the quantum and classical
solutions should be comparable. At the Planck scale ($x\equiv kz \sim 1$) the
wave functions could predict an intrinsically non-classical behaviour of
space-time.

\subsection{Induced 2d-gravity}

        On general grounds we can observe that the different behaviour
of the wave functions for large $x\equiv kz$ reflects appropriately the
classical solutions (\ref{dosxv}), (\ref{dosxvi}). If $\lambda^2 <0$
the universe starts with zero size ($t=-\infty$), expands to a maximum
radius $a_{max} = {1\over2\pi} {|r|\over\lambda}$,
and then contracts to zero size ($t=+\infty$) (the proper time
between $t=-\infty$ and $t=+\infty$ is finite). The wave functions
(\ref{ccuatrovid}) reflects this fact through the exponential decay
of the modified Hankel functions ${\cal K}_\nu$ for large $x$.
We can also understand the absence of the classical parabolic solution
$r=0$ (i.e. $\pi_\Phi = 0$, see (\ref{btresxiv})) due to absence of
normalizable solutions for $q=0$ (i.e. $\nu = \pm{1\over2}$).
For $\lambda^2 >0$, the classical parabolic solution does exist and their
quantum counterpart corresponds to the unique normalizable solution
with $q=0$, i.e. $<{\cal J}_{1\over2}>$.

        Now we wish to understand some quantitative aspects of the quantum
behaviour, as the $x^{-{1\over2}}$ decay of the wave functions for large
$x$ and positive cosmological constant. To this end we shall compare
the quantum probability distribution defined by the wave functions
($x^{-1} |\Psi|^2$) with the ``classical'' one ${\cal P}$ defined by
random sampling in the time domain
\be
{\cal P} \Delta x \approx \Delta{\cal T} \, .
\label{dcincoi}
\ee
To properly establish such a comparison we have to make a choice of
time to define ${\cal P}$ in (\ref{dcincoi}). As we have already
discussed, the natural time variable for the induced 2d-gravity is
the dilaton $\Phi$.

        For $\lambda^2 > 0$ the classical solutions can be rewritten as
\be
a(\Phi) = {|r|\over\lambda\pi} e^{-{(\Phi-\Phi_0)\over2}}
          ( 1 - e^{(\Phi-\Phi_0)\over2} )^{1\over2} \, ,
\label{dcincoii}
\ee
where
\be
e^{\Phi_0} = {1\over4} {\alpha\over(\sinh{r\over4\pi})^2} \, .
\label{dcincoiii}
\ee
The classical probability (${\cal P} \approx \left|{d\Phi\over da}\right|$)
is given by ($x_0 \neq 0$)
\be
{\cal P}^{\pm}_{\lambda^2 > 0} (x) = {1\over x_0}
        \left| { {x\over x_0} \over 1 + \left({x\over x_0}\right)^2 \pm
                                    \left( 1 + \left({x\over x_0}\right)^2
                                    \right)^{1\over2} }
        \right|
\, ,
\label{dcincoiv}
\ee
where $x_0 = {2|r|\over\pi}$ and the signs $\pm$ correspond to the
double possibility of
having an expanding or contracting dilaton field.
We have normalized (\ref{dcincoiv})
(with the positive sign) using the same
regularization procedure we have already introduced in section 4.

        The asymptotic behaviour of
${\cal P}^{\pm}_{\lambda^2>0}
\left( {\mathop{\sim}\limits_{{\scriptscriptstyle{x\rightarrow\infty}}}}
x^{-1} \right)$
fairly reproduce the decay of the probability distribution $x^{-1}|\Psi|^2$
for large $x$.
For real order the wave functions $<{\cal J}_\nu>$ and $<{\cal N}_\nu>$ are
oscillating for large $x$
and the local average of the asymptotic form of the (normalized) probability
distribution is just $x^{-1}$.
In Fig. I
we have plotted the functions
${\cal P}^{\pm}_{\lambda^2>0} (x)$ and the probability distribution of
the elementary (delta function normalized) wave functions $<{\cal J}_\nu>$
and $<{\cal N}_\nu>$ for $\nu={1\over4}$ ($x_0=\sqrt{{3\over32}}$). For
purely imaginary order $\nu=i y$, the probability distributions corresponding
to states in ${\cal H}^{(+)}$ and ${\cal H}^{(-)}$ oscillate between
${\cal P}^-$ and ${\cal P}^+$.

\begin{figure}
	\centerline{\psfig{figure=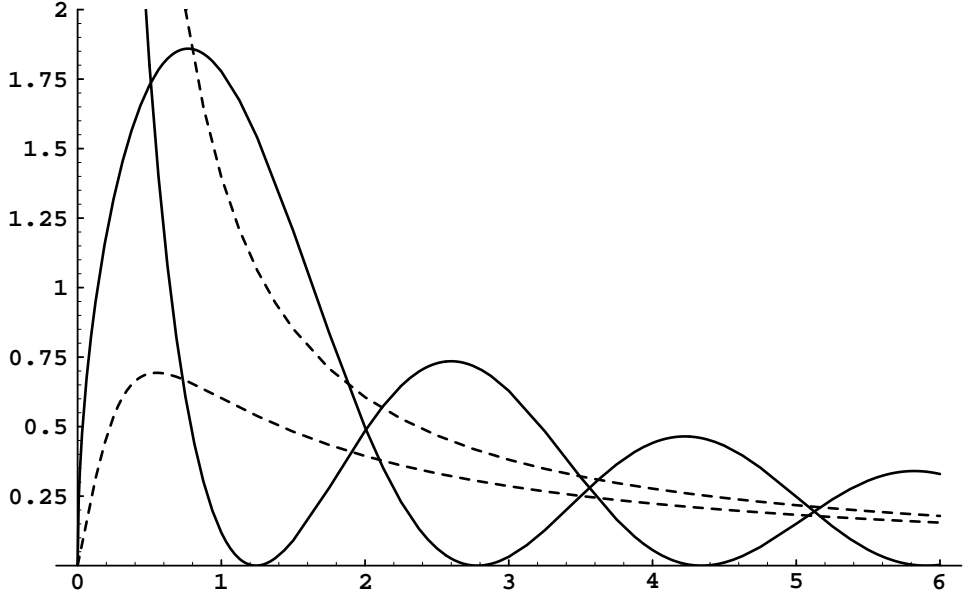,width=5in}}
\caption{Induced 2d-gravity. Probability distribution of
	the elementary (normalized) wave functions $<{\cal J}_\nu(x)>$
	and $<{\cal N}_\nu(x)>$ for $\nu={1\over4}$.
	The dotted lines are the functions
	${\cal P}^{\pm}_{\lambda^2>0} (x)$.}
\label{fig:hiperbola}
\end{figure}

        For $x_0 \rightarrow 0$ both ${\cal P}^+_{\lambda^2>0}$ and
${\cal P}^-_{\lambda^2>0}$ go to the (non normalizable) classical
distribution ${\cal P}_{\lambda^2 > 0} \propto {1\over x}$ but at the
quantum level only the solution $<{\cal J}_{1\over2}>$ is normalizable.

\subsection{CGHS-model}

        The comparison between the quantum and classical solutions
for the CGHS-model can also be traced along the lines of the induced
gravity model. For the CGHS-model the modified Bessel functions describe
the quantum states for $\lambda^2>0$ instead of $\lambda^2<0$, as for
the induced gravity. However the argument of the ${\cal K}_\nu(x)$
functions is now $x\equiv kA = k a e^{-2\Phi}$. If we consider the
classical probability distribution ${\cal P}_{\lambda^2<0} (x)$
defined in terms of the internal time variable $T=2\Phi + \ln a$ we
shall see that the quantum solutions also reflect appropriately their
classical behaviour.

        For $\lambda^2 > 0$ the classical solutions can be rewritten as
\be
A e^T - \left({r\over2\pi}\right)^2 {M\over|\lambda|^3} {1\over\sqrt{A}}
e^{T\over2} + \left({r\over2\pi}\right)^2 {1\over\lambda^2} = 0 \, ,
\label{dcincov}
\ee
and the classical probability is
\be
{\cal P}_{\lambda^2>0} (x) =
        {2\over x_{max}} \left| {1\over x} \left(
                {2\over\sqrt{1-\left({x\over x_{max}}\right)^2}} - 1 \right)
                \right|
        +
        {2\over 3 x_{max}} \left| {1\over x} \left(
                {2\over\sqrt{1-\left({x\over x_{max}}\right)^2}} + 1 \right)
                \right|
        \, ,
\label{dcincovi}
\ee
where $x_{max}=k A_{max}=k{M r\over 4\pi\lambda^2} = 2 {\pi_T\over\hbar}$.

        The expression (\ref{dcincovi}) has been normalized with the
regularization procedure of section 4. In
Fig. II
we have plotted ${\cal P}_{\lambda^2>0} (x)$ and the quantum distribution
$|{\cal K}_\nu (x)|^2$ for $x_{max}=16$ and
$\nu={1\over2}\sqrt{1024-1} i$.

\begin{figure}
	\centerline{\psfig{figure=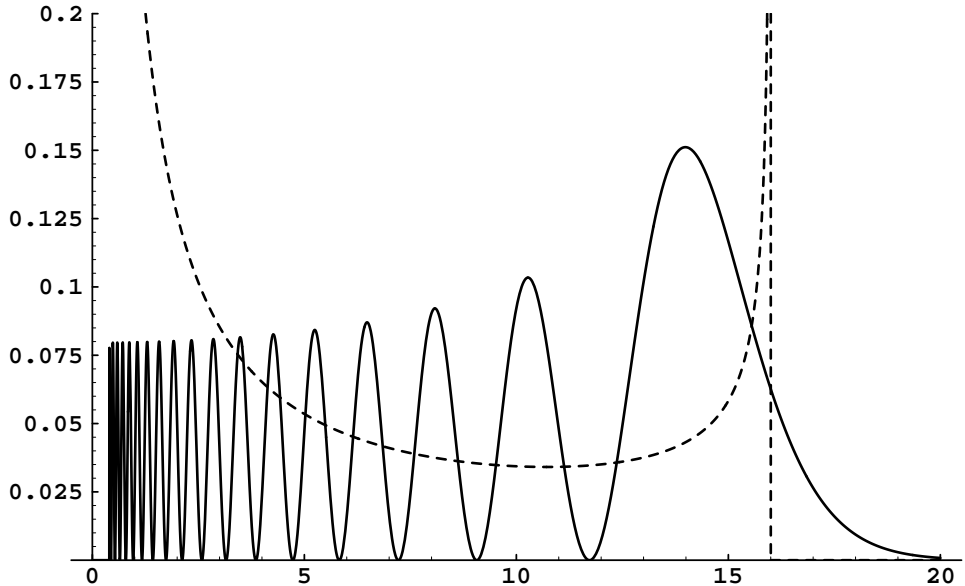,width=5in}}
\caption{CGHS-model. Quantum distribution %
	$|{\cal K}_{\nu} (x)|^2$ for
	$x_{max}=16$ and %
	${\nu}={1\over2}\protect\sqrt{1023}i$. %
	The dotted line is %
	${\cal P}_{\lambda^{2}>0} (x)$.}
\label{fig:acotado}
\end{figure}

        In the region $x\in]0,x_{max}[$ the wave function is essentially
oscillatory and it thus predicts classical behaviour.
In the classically forbidden region $x>x_{max}$ the wave function decays
exponentially. However, for $\left({x_{max}\over2}\right)^2 < 16$,
the order $\nu$ is real and the probability distribution is peaked around
the origin $x=0$ without any oscillatory (classical) region.
So that, the initial point $\nu=0$ separates two different phases of
the quantum solutions.
We must also note that the absence of normalizable states for $q=0$ (i.e.
$\nu=\pm{1\over2}$) reflects the absence of classical solutions with
vanishing $\pi_T$($={M r\over2\pi\lambda}$) (i.e. solutions with $M=0$).
A similar discussion can be given for the induced gravity (if $\lambda^2<0$)
and yield to the same kind of results.

	We can also consider the case of
negative cosmological constant. The discussion also parallels the
corresponding one of the induced gravity. We only want to mention that the
quantum state $<{\cal J}_{1\over2}>$ ($\nu={1\over2}$ and therefore $q=0$)
find its classical counterpart in the solution for which
$\pi_T={M r\over2\pi\lambda}$ vanishes. This singles out the solution
$M=0$, i.e. the linear vacuum dilaton (the analogue of the classical
parabolic solution of the induced gravity).

\subsection{Ground state}

        There are several proposals in the literature \cite{Hartle,Vilenkin}
to single out the wave function representing the ground state of the
gravitational field. Next we want to put forward an acceptable proposal
for the ground state of the models considered in this paper.
In studying the hamiltonian form of both the induced gravity and the
CGHS-model we were led to a natural choice of the internal time variable
${\cal T}$ (${\cal T}=\Phi$ for the induced gravity and
${\cal T} = 2\Phi +\ln a$ for the CGHS-model). These choices yield to
time-independent effective hamiltonians ($-\pi_\Phi$ (\ref{btresxxii})
and $-\pi_T$ (\ref{ctresxxvii}), respectively) and also allow an acceptable
physical interpretation of the quantum wave functions.
Therefore, it could also
be quite natural to define the grounds states as the state of
minimum ``internal'' energy (i.e. the state of minimum
${\hat\pi}_{\cal T}$-eigenvalue in our cases). Observe that the quantum
probability distributions
are not modified by the change $q \rightarrow -q$. In fact,
this transformation is equivalent to the change $t \rightarrow -t$
(which transforms an expanding dilaton solution into a contracting one)
and, therefore,
we should require the ground state be defined as the state of
minimum $|{\hat\pi}_{\cal T}|$.
In consequence the ground state of both models ($\lambda^ 2>0$ for
the induced gravity and $\lambda^2<0$ for the CGHS-model) is represented
by $<{\cal J}_{1\over2}>$. If $\lambda^2>0$ ($\lambda^2<0$) there is no
ground state for the induced gravity (CGHS-model).

	It is interesting to observe
that this wave function vanishes at $x=0$ and it is the one
having the maximum rate at which the probability density tends to
zero. This is reasonably in agreement with the ``no-boundary''
proposal \cite{Hartle}.

	As we have already mentioned the classical counterpart
of the quantum solution $\nu={1\over2}$ is represented by the
parabolic solution and the linear vacuum dilaton. We would
like to note that the quantum equivalence of the induced gravity
and the CGHS-model through the transformation $a \rightarrow A$,
$\Phi \rightarrow T$ also applies to the classical counterpart
of the ground state. The parabolic solution can be converted
into the linear dilaton solutions and both solutions take the form
$a \propto e^{-{\Phi\over2}}$ ($A \propto e^{-T}$).

\section{Final comments}

	In this paper we have carried out a novel canonical
analysis of the induced 2d-gravity and the CGHS-model on a
spatially closed universe. In the canonical quantization of
generally covariant theories (and gauge theories in general)
one can either first quantize and then impose the constraints
or first solve the constraints and then quantize. In the
present work we have followed a mixed procedure. First, we
have classically solved the supermomentum constraint and
fixed the spatial coordinate as well. The time coordinate
has been partially fixed by introducing a generic (spatially homogeneous)
gauge. This means a partial identification of time before
quantization. The residual reparametrization of the time coordinate
is incorporated at the quantum level by the (reduced) Wheeler-DeWitt
equation. We have completed the identification of time after
quantization, and it has played a major role in the space-time
interpretation of the quantum wave functions.

	A non-trivial test of the consistence of the present
approach is the full agreement we have found between the
quantization obtained via the Wheeler-DeWitt equation and
the one predicted by the covariant phase-space. The Hilbert
spaces determined by the geometric quantization of the symplectic
manifolds of the corresponding spaces
of non-equivalent classical solutions are
equivalent to those spanned by the (normalizable) solutions
of the Wheeler-DeWitt equation. Furthermore,
the quantum equivalence between the induced 2d-gravity
and the CGHS-model finds its classical counterpart also in their
covariant phase-space, and suggests a gauge-theoretical
formulation of the induced 2d-gravity mimicking the Poincar\`e
extended formulation of the CGHS-model \cite{Cangemi}.

	Finally we want to point out that,
when the cosmological constant vanishes, the
wave functions coincide exactly with the small $x$ behaviour
of the wave functions in the general case.
Therefore, the wave functions are not normalizable and
the approach breaks down.
The role of the cosmological term is thus to modify the
large $x$ behaviour of the wave functions leading then to
normalizable states.

\section*{Acknowledgements}

	C. F. Talavera acknowledges to the {\it Generalitat Valenciana}
for a FPI fellowship. Both authors would like to thank M. Navarro
for useful discussions.

\end{document}